\documentclass[prl,amsfonts,twocolumn,showpacs,floatfix]{revtex4-1}
\usepackage{amsmath}
\usepackage{amsfonts}
\usepackage{amssymb}
\usepackage{nicefrac}
\usepackage{tabularx}
\usepackage{graphicx,color,epsfig,rotate}
  
\begin{document}
\title{Magnetic order induced by Fe substitution of Al site in the heavy fermion systems $\alpha$-YbAlB$_4$ and $\beta$-YbAlB$_4$}
\author{Kentaro Kuga$^1$, Gregory Morrison$^2$, LaRico Treadwell$^2$, Julia Y. Chan$^2$ and Satoru Nakatsuji$^1$ }
\email{satoru@issp.u-tokyo.ac.jp}
\affiliation{$^1$Institute for Solid State Physics (ISSP), University of Tokyo, Kashiwa 277-8581, Japan}
\affiliation{$^2$Department of Chemistry, Louisiana State University, Baton Rouge, Louisiana 70803, USA}
\date{\today}
\begin{abstract}
$\beta$-YbAlB$_4$ is a heavy fermion superconductor that exhibits a quantum criticality without tuning at zero field and under ambient pressure. We have succeeded in substituting Fe for Al in $\beta$-YbAlB$_4$ as well as the polymorphous compound $\alpha$-YbAlB$_4$, which in contrast has a heavy Fermi liquid ground state. Full structure determination by single crystal X-ray diffraction confirmed no change in crystal structure for both $\alpha$- and $\beta$-YbAlB$_4$, in addition to volume contraction with Fe substitution. Our measurements of the magnetization and specific heat indicate that both $\alpha$-YbAl$_{0.93}$Fe$_{0.07}$B$_4$ and $\beta$-YbAl$_{0.94}$Fe$_{0.06}$B$_4$ exhibit a magnetic order, most likely of a canted antiferromagnetic type, at $7\sim 9$ K. The increase in the entropy as well as the decrease in the antiferromagnetic Weiss temperature with the Fe substitution in both systems indicates that the chemical pressure due to the Fe substitution suppresses the Kondo temperature and induces the magnetism.
\end{abstract}
\maketitle

4$f$-electron based heavy fermion systems have provided prototypical systems to study interesting phenomena in the vicinity of quantum critical points, such as unconventional superconductivity and non-Fermi-liquid states \cite{Vojta,Monthoux,Gegenwart}. In particular, much attention has been recently paid to unconventional quantum critical materials, such as CeCu$_{6-x}$Au$_x$, YbRh$_2$Si$_2$, and $\beta$-YbAlB$_4$ \cite{Lohneysen,Trovarelli,Nakatsuji}, that exhibit novel types of quantum criticality beyond the standard spin-density-wave description \cite{Moriya,Millis}.

The Yb-based heavy fermion system, YbAlB$_4$, has two polymorphs with different structures: noncentrosymmetric $\alpha$-YbAlB$_4$ and centrosymmetric $\beta$-YbAlB$_4$ \cite{Nakatsuji,Macaluso}. $\beta$-YbAlB$_4$ is the first example of an Yb-based heavy fermion superconductor with the transition temperature $T_c$ of 80 mK \cite{Nakatsuji,Kuga}. Moreover, it is a unique example of a metal that exhibits a quantum criticality without tuning of any control parameters \cite{Nakatsuji,Matsumoto1}. Strong sensitivity of the non-Fermi-liquid behavior to the magnetic field, in particular, the $T/B$ scaling of the magnetization indicate that the quantum critical point of $\beta$-YbAlB$_4$ should be located exactly at zero field under ambient pressure \cite{Matsumoto1}. On the other hand, low temperature behavior of $\alpha$-YbAlB$_4$ is well fit to a Fermi liquid type description and forms a heavy Fermi liquid state with the specific heat coefficient $\gamma \sim$ 130 mJ/mol K$^2$ below around $T^* \sim 8$ K. 

\begin{table*}
\begin{center}
\caption{Crystallographic Data for $\alpha$-YbAl$_{0.93}$Fe$_{0.07}$B$_{4}$, $\beta$-YbAl$_{0.97}$Fe$_{0.03}$B$_{4}$ and $\beta$-YbAl$_{0.95}$Fe$_{0.05}$B$_{4}$ at 295(3) and 100(1) K}
\centering
\newcolumntype{X}{>{\centering\arraybackslash}p{22mm}} 
\begin{tabular}{p{38mm}XXXXXX}
\\
  \hline
  \hline
\multicolumn{1}{l}{Formula} & \multicolumn{2}{c}{\textbf{$\alpha$-YbAl$_{0.93}$Fe$_{0.07}$B$_{4}$}} & \multicolumn{2}{c}{\textbf{$\beta$-YbAl$_{0.97}$Fe$_{0.03}$B$_{4}$}} & \multicolumn{2}{c}{\textbf{$\beta$-YbAl$_{0.95}$Fe$_{0.05}$B$_{4}$}} \\
  \hline
  \hline
 Temperature (K) & 295(3) & 100(1) & 295(3) & 100(1) & 295(3) & 100(1) \\
Space group & P\textit{bam} & P\textit{bam} & C\textit{mmm} & C\textit{mmm} & C\textit{mmm} & C\textit{mmm} \\
$a$ (\AA{}) & 5.9184(6) & 5.9167(6) & 7.3060(9) & 7.3010(12) & 7.3020(12) & 7.3010(12) \\
$b$ (\AA{}) & 11.4645(15) & 11.4602(15) & 9.3180(12) & 9.3130(12) & 9.3180(12) & 9.3130(12) \\
$c$ (\AA{}) & 3.4832(6) & 3.4780(4) & 3.4970(3) & 3.4890(3) & 3.4920(6) & 3.4850(6) \\
$V$ (\AA{}$^3$) & 236.34(6) & 235.83(4) & 238.07(5) & 237.23(5) & 237.60(6) & 236.96(6) \\
$Z$ & 4 & 4 & 4 & 4 & 4 & 4 \\
\multicolumn{1}{l}{Crystal dimensions (mm$^{3}$)} & \multicolumn{2}{c}{0.07 $\times$ 0.13 $\times$ 0.17} & \multicolumn{2}{c}{0.01 $\times$ 0.07 $\times$ 0.07} & \multicolumn{2}{c}{0.01 $\times$ 0.07 $\times$ 0.07} \\
Density (g cm$^{-3}$) & 6.894 & 6.904 & 6.787 & 6.811 & 6.801 & 6.819 \\
$\theta$ Range ($^\circ$) & 3.55-31.01 & 3.56-30.97 & 3.54-30.92 & 3.54-30.99 & 3.54-30.96 & 3.55-30.90 \\
\textit{$\mu$} (mm$^{-1}$) & 39.918 & 39.972 & 39.242 & 39.380 & 39.320 & 39.425 \\
\\
\multicolumn{2}{l}{\textit{Data Collection and Refinement}} \\
Collected reflections & 3650 & 4178 & 2160 & 2252 & 2251 & 2598 \\
Unique reflections & 433 & 426 & 244 & 244 & 244 & 243 \\
$R_{\rm{int}}$ & 0.0179 & 0.0155 & 0.0133 & 0.0128 & 0.0234 & 0.0203 \\
\multicolumn{1}{l}{\textit{h}} & \multicolumn{2}{c}{-8 $\leq$ \textit{h} $\leq$ 8} & \multicolumn{2}{c}{-10 $\leq$ \textit{h} $\leq$ 10} & \multicolumn{2}{c}{-10 $\leq$ \textit{h} $\leq$ 10} \\
\multicolumn{1}{l}{\textit{k}} & \multicolumn{2}{c}{-16 $\leq$ \textit{k} $\leq$ 16} & \multicolumn{2}{c}{-13 $\leq$ \textit{k} $\leq$ 13} & \multicolumn{2}{c}{-13 $\leq$ \textit{k} $\leq$ 13} \\
\multicolumn{1}{l}{\textit{l}} & \multicolumn{1}{c}{-5 $\leq$ \textit{l} $\leq$ 5} & \multicolumn{1}{c}{-4 $\leq$ \textit{l} $\leq$ 5} & \multicolumn{2}{c}{-5 $\leq$ \textit{l} $\leq$ 5} & \multicolumn{2}{c}{-4 $\leq$ \textit{l} $\leq$ 5} \\
$\Delta\rho_{\rm{max}}$ (e \AA{}$^{-3}$) & 3.068 & 2.999 & 2.611 & 2.474 & 5.223 & 4.764 \\
$\Delta\rho_{\rm{min}}$ (e \AA{}$^{-3}$) & -1.862 & -1.768 & -2.607 & -2.758 & -4.680 & -3.777 \\
GoF & 1.149 & 1.294 & 1.149 & 1.159 & 1.183 & 1.159 \\
Extinction coefficient & 0.0121(10) & 0.0119(10) & 0.0166(11) & 0.0144(10) & 0.018(3) & 0.014(3) \\
$^aR_1(\textit{F})$ for $\textit{F}_{\rm{o}}{}^2 > 2\sigma(\textit{F}_{\rm{o}}{}^2)$ & 0.0209 & 0.0237 & 0.0177 & 0.0186 & 0.0388 & 0.0424 \\
$^bR_{\rm{w}}(\textit{F}_{\rm{o}}{}^2)$ & 0.0553 & 0.0603 & 0.0459 & 0.0462 & 0.1016 & 0.1116 \\ [0.5ex]
\hline
\multicolumn{7}{p{17.7cm}}{\scriptsize $^aR_1 = \sum||F_{\rm{o}}|-|F_{\rm{c}}||/\sum|F_{\rm{o}}|$} \\
\multicolumn{7}{p{17.7cm}}{\scriptsize$^b$\textit{w}$R_1 = [\sum\textit{w}(F_{\rm{o}}{}^2 - F_{\rm{c}}{}^2)^2]^{1/2}; P = (F_{\rm{o}}{}^2 + 2F_{\rm{c}}{}^2)/3; \textit{w} = 1/[\sigma^2(F_{\rm{o}}{}^2) + (0.0261P)^2 + 3.2341P], \textit{w} = 1/[\sigma^2(F_{\rm{o}}{}^2) + (0.0258P)^2 + 4.5691P], \textit{w} = 1/[\sigma^2(F_{\rm{o}}{}^2) + (0.0334P)^2 + 0.8817P], \textit{w} = 1/[\sigma^2(F_{\rm{o}}{}^2) + (0.0318P)^2 + 2.0681P], \textit{w} = 1/[\sigma^2(F_{\rm{o}}{}^2) + (0.0852P)^2], $ and $ \textit{w} = 1/[\sigma^2(F_{\rm{o}}{}^2) + (0.0968P)^2] $ for $\alpha$-YbAl$_{0.93}$Fe$_{0.07}$B$_{4}$, $\beta$-YbAl$_{0.97}$Fe$_{0.03}$B$_{4}$ and $\beta$-YbAl$_{0.95}$Fe$_{0.05}$B$_{4}$ at 295 K and 100 K, respectively} \\
\end{tabular}
\end{center}
\end{table*}

Up to date, all the 4$f$-electron based quantum critical materials have the valence close to the integer, providing good evidence that these systems can be well described by the Kondo lattice model \cite{Vojta,Gegenwart}. In sharp contrast, both polymorphs of YbAlB$_4$ are found to be valence fluctuating systems with a strongly intermediate valence, such as Yb$^{2.73+}$ for $\alpha$-YbAlB$_4$ and Yb$^{2.75+}$ for $\beta$-YbAlB$_4$ at 20 K \cite{Okawa}. Strong hybridization has been confirmed by the itinerant $f$-electron character found in the quantum oscillation study of the Fermi surface of $\beta$-YbAlB$_4$ \cite{eoinPRL}. 

Interestingly, however, both $\alpha$- and $\beta$-YbAlB$_4$ show Kondo lattice behaviors in the low energy thermodynamics at low temperatures below the characteristic temperature $T^*$ of 8 K \cite{Matsumoto1,Matsumoto2} and exhibit local moment behavior of Yb$^{3+}$ state in the temperature dependence of the susceptibility and in the electron spin resonance spectra, particularly for the $\beta$ phase below $T^*$ \cite{Holanda}. Recent Hall resistivity measurements for $\beta$-YbAlB$_4$ revealed a peak at $\sim 40$ K, indicating that the coherence appears at a much lower temperature than expected for a strongly intermediate valence state\cite{eoinPRL2}.

In order to understand the origin of the quantum criticality found under ambient condition for $\beta$-YbAlB$_4$ as well as the unusual Kondo lattice behaviour in the valence fluctuation states found in both $\alpha$ and $\beta$ phases, it is highly important to reveal the nature of magnetic instability existing nearby the quantum criticality and heavy fermion state. Here, we report our discovery that Fe substitution for Al site causes antiferromagnetic order in both phases of YbAlB$_4$. Adopting the chemical substitution as a control parameter, we have succeeded in substituting Fe for Al, and discovered that a small Fe substitution of around $5 \sim 7$\% is enough to induce an antiferromagnetic order at $7 \sim 9$ K in both $\alpha$- and $\beta$-phases of YbAlB$_4$. Our high precision determination of the crystal structure indicates no change in the crystal structure by Fe substitution or by temperature sweep at least down to 100 K. The results of the low temperature susceptibility and specific heat measurements show much stronger temperature dependence than in the pure YbAlB$_4$, indicating that the chemical pressure induced by Fe substitution reduces the Kondo temperature and thereby induces magnetic order.

We have succeeded in growing single crystals of $\alpha$-YbAl$_{0.93}$Fe$_{0.07}$B$_4$, $\beta$-YbAl$_{0.97}$Fe$_{0.03}$B$_4$, $\beta$-YbAl$_{0.95}$Fe$_{0.05}$B$_4$, and $\beta$-YbAl$_{0.94}$Fe$_{0.06}$B$_4$ using the Al flux growth technique. We have also grown single crystals of a Lu-analog, $\alpha$-LuAl$_{0.79}$Fe$_{0.21}$B$_4$, and $\beta$-LuAl$_{0.96}$Fe$_{0.04}$B$_4$. The temperature and field dependence of the magnetization was measured using the commercial SQUID magnetometer (MPMS, Quantum Design). The temperature dependence of the specific heat $C_P$ was measured using a relaxation method. The entropy was estimated by integrating $C_P/T$ over temperature from the lowest temperature 0.4 K of the measurement. The Fe concentration for $\alpha$-YbAl$_{0.93}$Fe$_{0.07}$B$_4$ was estimated using inductively coupled plasma (ICP) spectroscopy within the resolution of 0.3\%, and for $\beta$-YbAl$_{1-x}$Fe$_{x}$B$_{4}$, $\alpha$-LuAl$_{0.79}$Fe$_{0.21}$B$_4$, and $\beta$-LuAl$_{0.96}$Fe$_{0.04}$B$_4$ by energy dispersive X-ray analysis (EDX) within the resolution of 3\%. We also utilized the ICP method for some of $\beta$-YbAl$_{1-x}$Fe$_{x}$B$_{4}$ samples and confirmed the Fe concentration within 1 \% difference from the EDX results. For example, a single crystal of $\beta$-YbAl$_{1-x}$Fe$_{x}$B$_{4}$ is found to have $x = 0.03$ by EDX, and $x = 0.02$ by ICP, respectively. Hereafter, we use $x$(Fe) determined by EDX method for all the samples of $\beta$-YbAl$_{1-x}$Fe$_{x}$B$_{4}$ and ICP results for $\alpha$-YbAl$_{1-x}$Fe$_{x}$B$_{4}$.

\begin{table*}
\caption{Atomic Coordinates and Displacement Parameters for $\alpha$-Yb$_{1}$Al$_{1-x}$Fe$_{x}$B$_{4} ~(x = 0.07)$ at 295(3) and 100(1) K}
\centering
\newcolumntype{Y}{>{\centering\arraybackslash}p{22mm}} 
\begin{tabular}{YYYYYYY}
\\ [-2ex]
\hline
\\ [-2ex]
Atom & Wyckoff site & \textit{x} & \textit{y} & \textit{z} & \textit{U}$_{\rm{eq}}$ (\AA{}$^2)^\textit{a}$ & Occ. \\  [0.5ex]
\hline
\\ [-2ex]
\multicolumn{2}{l}{\textbf{295(3) K}}\\
Yb1 & 4\textit{g} & 0.12859(5) & 0.15052(3) & 0 & 0.00344(17) & 1 \\
Al1 & 4\textit{g} & 0.1365(4) & 0.4109(2) & 0 & 0.0042(8) & 0.929(15) \\
Fe1 & 4\textit{g} & 0.1365(4) & 0.4109(2) & 0 & 0.0042(8) & 0.071(15) \\
B1 & 4\textit{h} & 0.2921(16) & 0.3135(8) & $\nicefrac{1}{2}$ & 0.0056(16) & 1 \\
B2 & 4\textit{h} & 0.3654(15) & 0.4695(8) & $\nicefrac{1}{2}$ & 0.0054(16) & 1 \\
B3 & 4\textit{h} & 0.3850(16) & 0.0479(8) & $\nicefrac{1}{2}$ & 0.0063(15) & 1 \\
B4 & 4\textit{h} & 0.4725(16) & 0.1939(8) & $\nicefrac{1}{2}$ & 0.0042(16) & 1 \\
\\
\multicolumn{2}{l}{\textbf{100(1) K}}\\
Yb1 & 4\textit{g} & 0.12862(6) & 0.15052(3) & 0 & 0.00166(19) & 1 \\
Al1 & 4\textit{g} & 0.1361(4) & 0.4109(2) & 0 & 0.0023(9) & 0.935(17) \\
Fe1 & 4\textit{g} & 0.1361(4) & 0.4109(2) & 0 & 0.0023(9) & 0.065(17) \\
B1 & 4\textit{h} & 0.2915(19) & 0.3138(9) & $\nicefrac{1}{2}$ & 0.0047(18) & 1 \\
B2 & 4\textit{h} & 0.3651(17) & 0.4686(8) & $\nicefrac{1}{2}$ & 0.0032(17) & 1 \\
B3 & 4\textit{h} & 0.3875(17) & 0.0482(9) & $\nicefrac{1}{2}$ & 0.0023(16) & 1 \\
B4 & 4\textit{h} & 0.4751(18) & 0.1929(9) & $\nicefrac{1}{2}$ & 0.0037(18) & 1 \\
\hline
\multicolumn{7}{l}{\scriptsize $^a$\textit{U}$_{\rm{eq}}$ is defined as one-third of the trace of the orthogonalized \textit{U}$_{\textit{ij}}$ tensor.} \\
\end{tabular}

\caption{Atomic Coordinates and Displacement Parameters for $\beta$-YbAl$_{1-x}$Fe$_{x}$B$_{4} ~({x = 0.03}^\textit{a}$) at 295(3) and 100(1) K}
\centering
\begin{tabular}{YYYYYYY}
\\ [-2ex]
\hline
\\ [-2ex]
Atom & Wyckoff site & \textit{x} & \textit{y} & \textit{z} & \textit{U}$_{\rm{eq}}$ (\AA{}$^2)^\textit{b}$ & Occ. \\  [0.5ex]
\hline
\\ [-2ex]
\multicolumn{2}{l}{\textbf{295(3) K}}\\
Yb1 & 4\textit{i} & 0 & 0.30065(3) & 0 & 0.00388(19) & 1 \\
Al1 & 4\textit{g} & 0.1808(3) & 0 & 0 & 0.0042(8) & 0.987(15) \\
Fe1 & 4\textit{g} & 0.1808(3) & 0 & 0 & 0.0042(8) & 0.013(15) \\
B1 & 4\textit{h} & 0.1219(13) & $\nicefrac{1}{2}$ & $\nicefrac{1}{2}$ & 0.0050(13) & 1 \\
B2 & 8\textit{q} & 0.2226(8) & 0.1594(14) & $\nicefrac{1}{2}$ & 0.0061(10) & 1 \\
B3 & 4\textit{j} & 0 & 0.0922(9) & $\nicefrac{1}{2}$ & 0.0044(13) & 1 \\
\\
\multicolumn{2}{l}{\textbf{100(1) K}}\\
Yb1 & 4\textit{i} & 0 & 0.30066(3) & 0 & 0.00260(19) & 1 \\
Al1 & 4\textit{g} & 0.1810(3) & 0 & 0 & 0.0031(8) & 0.996(15) \\
Fe1 & 4\textit{g} & 0.1810(3) & 0 & 0 & 0.0031(8) & 0.004(15) \\
B1 & 4\textit{h} & 0.1215(13) & $\nicefrac{1}{2}$ & $\nicefrac{1}{2}$ & 0.0046(14) & 1 \\
B2 & 8\textit{q} & 0.2224(8) & 0.1596(8) & $\nicefrac{1}{2}$ & 0.0046(10) & 1 \\
B3 & 4\textit{j} & 0 & 0.0921(9) & $\nicefrac{1}{2}$ & 0.0043(14) & 1 \\
\hline
\multicolumn{7}{l}{\scriptsize $^a$ $x = 0.03(2)$ is estimated by EDX method, while ICP yields $x = 0.02(1)$ with better accuracy, close to thea above Fe occupancy}\\
\multicolumn{7}{l}{\scriptsize obtained from the structural analysis.}\\
\multicolumn{7}{l}{\scriptsize $^b$\textit{U}$_{\rm{eq}}$ is defined as one-third of the trace of the orthogonalized \textit{U}$_{\textit{ij}}$ tensor.} \\
\end{tabular}

\caption{Atomic Coordinates and Displacement Parameters for $\beta$-YbAl$_{1-x}$Fe$_{x}$B$_{4} ~(x = 0.05)$ at 295(3) and 100(1) K}
\centering
\begin{tabular}{YYYYYYY}
\\ [-2ex]
\hline
\\ [-2ex]
Atom & Wyckoff site & \textit{x} & \textit{y} & \textit{z} & \textit{U}$_{\rm{eq}}$ (\AA{}$^2)^\textit{a}$ & Occ. \\  [0.5ex]
\hline
\\ [-2ex]
\multicolumn{2}{l}{\textbf{295(3) K}}\\
Yb1 & 4\textit{i} & 0 & 0.30070(4) & 0 & 0.0054(4) & 1 \\
Al1 & 4\textit{g} & 0.1799(6) & 0 & 0 & 0.0049(8) & 1 \\
B1 & 4\textit{h} & 0.122(3) & $\nicefrac{1}{2}$ & $\nicefrac{1}{2}$ & 0.005(2) & 1 \\
B2 & 8\textit{q} & 0.2225(12) & 0.1607(15) & $\nicefrac{1}{2}$ & 0.0067(17) & 1 \\
B3 & 4\textit{j} & 0 & 0.0917(15) & $\nicefrac{1}{2}$ & 0.006(2) & 1 \\
\\
\multicolumn{2}{l}{\textbf{100(1) K}}\\
Yb1 & 4\textit{i} & 0 & 0.30072(5) & 0 & 0.0042(4) & 1 \\
Al1 & 4\textit{g} & 0.1802(6) & 0 & 0 & 0.0031(8) & 1 \\
B1 & 4\textit{h} & 0.124(3) & $\nicefrac{1}{2}$ & $\nicefrac{1}{2}$ & 0.009(3) & 1 \\
B2 & 8\textit{q} & 0.2227(15) & 0.1598(18) & $\nicefrac{1}{2}$ & 0.009(2) & 1 \\
B3 & 4\textit{j} & 0 & 0.0913(12) & $\nicefrac{1}{2}$ & 0.004(3) & 1 \\
\hline
\multicolumn{7}{l}{\scriptsize $^a$\textit{U}$_{\rm{eq}}$ is defined as one-third of the trace of the orthogonalized \textit{U}$_{\textit{ij}}$ tensor.} \\
\end{tabular}
\end{table*}

The single crystal X-ray diffraction experiments were collected using a Nonius Kappa CCD diffractometer equipped with a Mo K$_{\alpha}$ source ($\lambda$ = 0.711 \AA) at room temperature, 295(3) K, and at 100(1) K. Direct methods using SIR97 \cite{Altomare} was performed to obtain an initial structural model which was then refined using SHELXL-97 \cite{Sheldrick}. Crystallographic data and atomic coordinates for $\alpha$-YbAl$_{1-x}$Fe$_x$B$_4$ ($x$ = 0.07) and $\beta$-YbAl$_{1-x}$Fe$_x$B$_4$ ($x$ = 0.03, 0.05) can be found in Tables 1-4. In comparison with the undoped analogue, the results clearly show a volume contraction of $\sim 0.8$\% for $\alpha$-YbAl$_{0.93}$Fe$_{0.07}$B$_4$,  $\sim 0.02$\% for $\beta$-YbAl$_{0.97}$Fe$_{0.03}$B$_4$, and of $\sim 0.2$\% for $\beta$-YbAl$_{0.95}$Fe$_{0.05}$B$_4$, indicating that the Fe doping applies a chemical pressure.

Table 2 shows the atomic coordinates for $\alpha$-YbAl$_{1-x}$Fe$_x$B$_4$ ($x$ = 0.07).  When the Fe substitution was not accounted for, the Al site (4$g$) had an anomalously small atomic displacement parameter as compared to the undoped $\alpha$-YbAlB$_4$. Therefore, the Fe was partially substituted on the Al site and the occupancies of the two elements were freely refined.  This resulted in a mixed occupancy of 7.1(15)\% Fe and 92.9(15)\% Al at 295(3) K and is in good agreement with the composition as obtained from the elemental analysis using the ICP method.  Furthermore, no evidence for a structural transition was observed with substitution, and likewise, no structural transition was observed for any Fe concentration upon cooling from room temperature down to 100(1) K.

Tables 3 and 4 provide atomic coordinates for $\beta$-YbAl$_{1-x}$Fe$_{x}$B$_{4} ~(x = 0.03, 0.05)$. When the Fe substitution was not included in the model, the Al site (4\textit{g}) had a similar atomic displacement parameter to the Yb site (4\textit{i}) suggesting that the Fe occupies the Al site. For $\beta$-YbAl$_{0.97}$Fe$_{0.03}$B$_4$, Fe was partially substituted onto the Al site and the occupancies were freely refined.  The resulting site occupancy was 1.3(15)\% Fe and 98.7(15)\% Al at 295(3) K.  This is in agreement with the ICP data which indicated the stoichiometry to be $\beta$-YbAl$_{0.98}$Fe$_{0.02}$B$_4$.  While the atomic displacement parameters also suggested that the Fe occupied the Al site in $\beta$-YbAl$_{0.95}$Fe$_{0.05}$B$_4$, no Fe could be refined onto this site, or any other site, for the model.  The inability to model the Fe doping in $\beta$-YbAl$_{0.95}$Fe$_{0.05}$B$_4$ can be attributed to the lower quality diffraction data for this analogue compared to the other analogues.  $\beta$-YbAl$_{0.95}$Fe$_{0.05}$B$_4$ grew as thin plates whereas $\beta$-YbAl$_{0.97}$Fe$_{0.03}$B$_4$ grew as thick plates and $\alpha$-YbAl$_{0.93}$Fe$_{0.07}$B$_4$ grew as rods.  The thinner plates for $\beta$-YbAl$_{0.95}$Fe$_{0.05}$B$_4$ led to lower quality X-ray diffraction data which is apparent in both the increased R$_{1}$ and residual electron densities for this analogue.  Due to the increased $\Delta\rho_{min/max}$, the Fe substitution could not be modeled.  As with $\alpha$-YbAl$_{1-x}$Fe$_x$B$_4$, no structural transition was observed in $\beta$-YbAl$_{1-x}$Fe$_x$B$_4$ upon doping or cooling from 298 K down to 100 K.

Single crystal X-ray diffraction data was also collected on a sample of $\beta$-YbAl$_{1-x}$Fe$_{x}$B$_{4} ~(x = 0.06)$. Although the mosaicity of the single crystal was suitable for data collection at 295(3) K, upon cooling in 50 K intervals down to 100(1) K, a continuous decrease in crystal quality, indicated by increased $\chi^2$s and mosaicity, was observed. For example, the mosaicity of the crystal increased from 0.45 degrees at 295(3) K to 0.87 degrees at 100(1) K.  When the crystal was warmed back to room temperature, the crystal quality returned to its original state. Diffraction data of $\beta$-YbAl$_{1-x}$Fe$_{x}$B$_{4} ~(x = 0.06)$ was collected at both 295(3) K and 100(1) K and no evidence for a structural transition was observed. In $\alpha$-YbAl$_{1-x}$Fe$_x$B$_4$ ($x$ = 0.07) and $\beta$-YbAl$_{1-x}$Fe$_{x}$B$_{4} ~(x = 0.03, 0.05)$, on the other hand, the degradation of crystal quality on cooling was not observed. Our synthesis experiments suggest that the concentration $x = 0.06$ is close to the edge of the stability of the Fe doped $\beta$-phase, and this may be the origin of the increase in the mosaicity on cooling. The cause of this decrease in crystal quality is currently being explored and will be the subject of a future manuscript.

\begin{figure*}
\begin{minipage}{19pc}
\includegraphics[width=19pc]{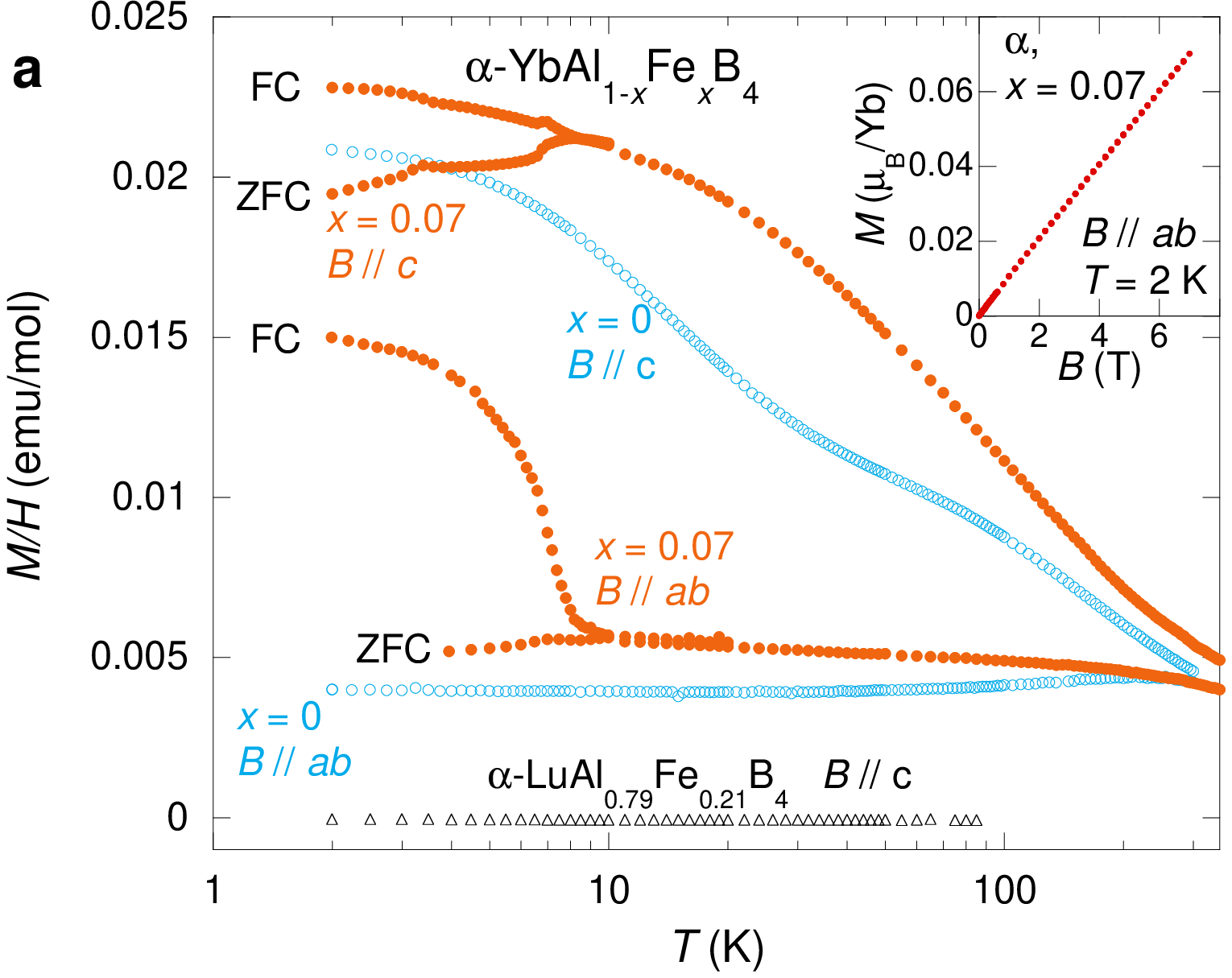}
\end{minipage}\hspace{1pc}%
\begin{minipage}{19pc}
\includegraphics[width=19pc]{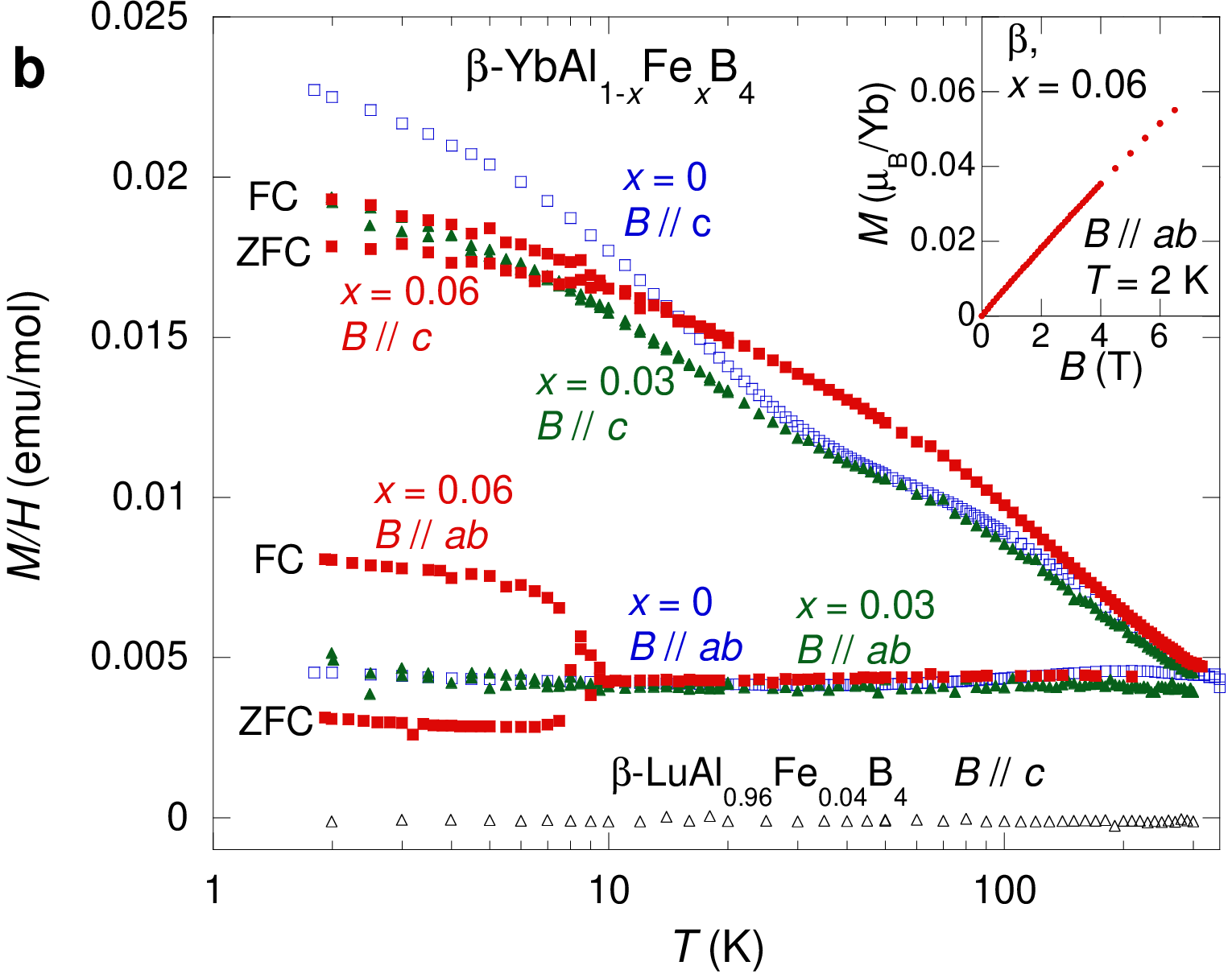}
\end{minipage}\hspace{2pc}%
\caption{(a) Temperature dependence of the $ab$-plane and $c$-axis susceptibility of $\alpha$-YbAlB$_4$ (open circle), $\alpha$-YbAl$_{0.93}$Fe$_{0.07}$B$_4$ (closed circle) measured under a field of 10 mT, and of $\alpha$-LuAl$_{0.79}$Fe$_{0.21}$B$_4$ (open triangle)  under 1 T. (b) Temperature dependence of the $ab$-plane and $c$-axis susceptibility of $\beta$-YbAlB$_4$ (open square), $\beta$-YbAl$_{0.97}$Fe$_{0.03}$B$_4$ (closed triangle) and $\beta$-YbAl$_{0.94}$Fe$_{0.06}$B$_4$ (closed square) measured under a field of 10 mT, and of $\beta$-LuAl$_{0.96}$Fe$_{0.04}$B$_4$ under 1 T. For all measurements, both zero-field-cooling and field-cooling sequences are employed. The inset of each panel shows the magnetic field dependence of the magnetization $M$ of $\alpha$-YbAl$_{0.93}$Fe$_{0.07}$B$_4$ and $\beta$-YbAl$_{0.94}$Fe$_{0.06}$B$_4$ under a field up to 7 T along $ab$ plane at 2 K.}
\end{figure*}

\begin{figure*}
\begin{minipage}{19pc}
\includegraphics[width=19pc]{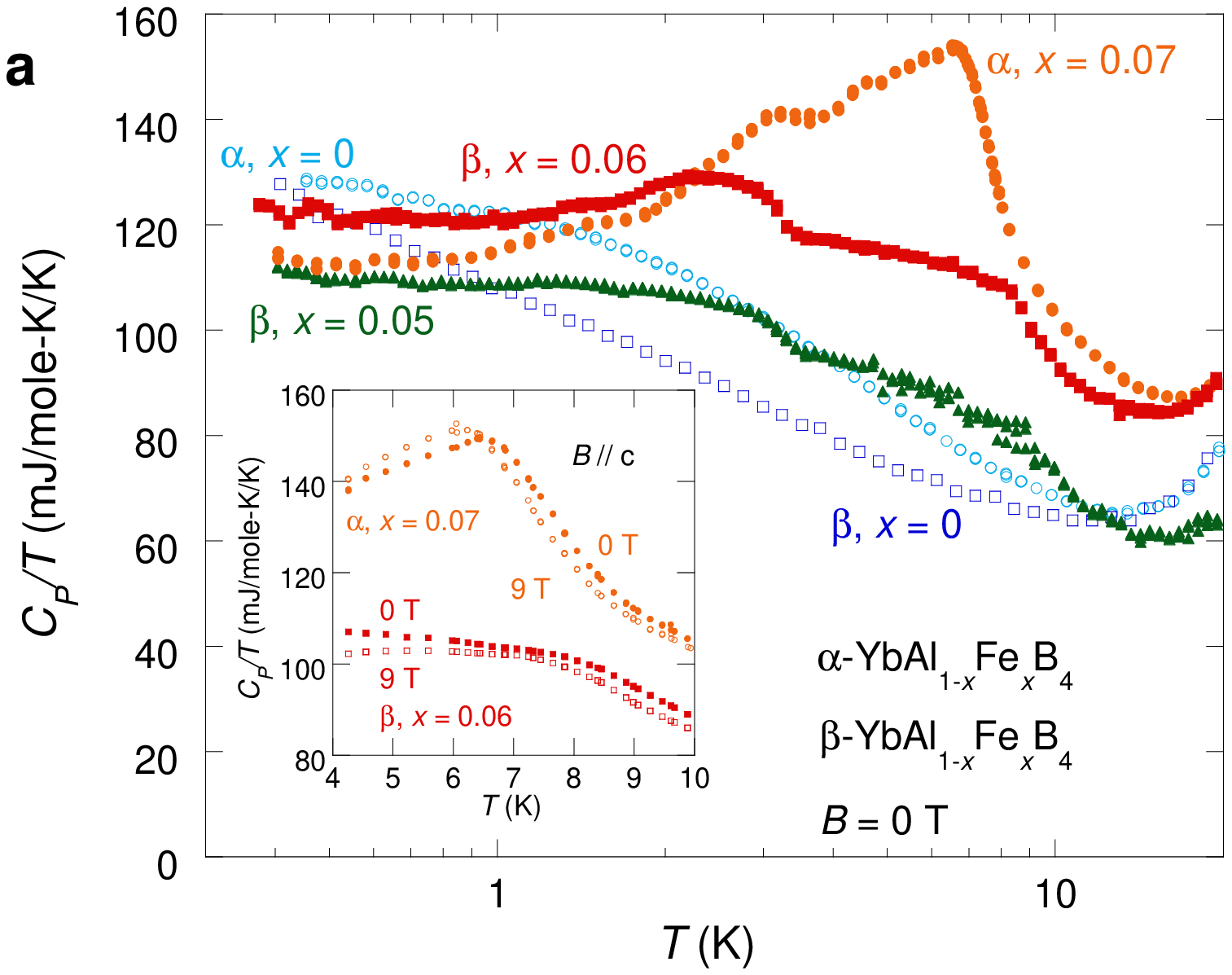}
\end{minipage}\hspace{1pc}%
\begin{minipage}{19pc}
\includegraphics[width=19pc]{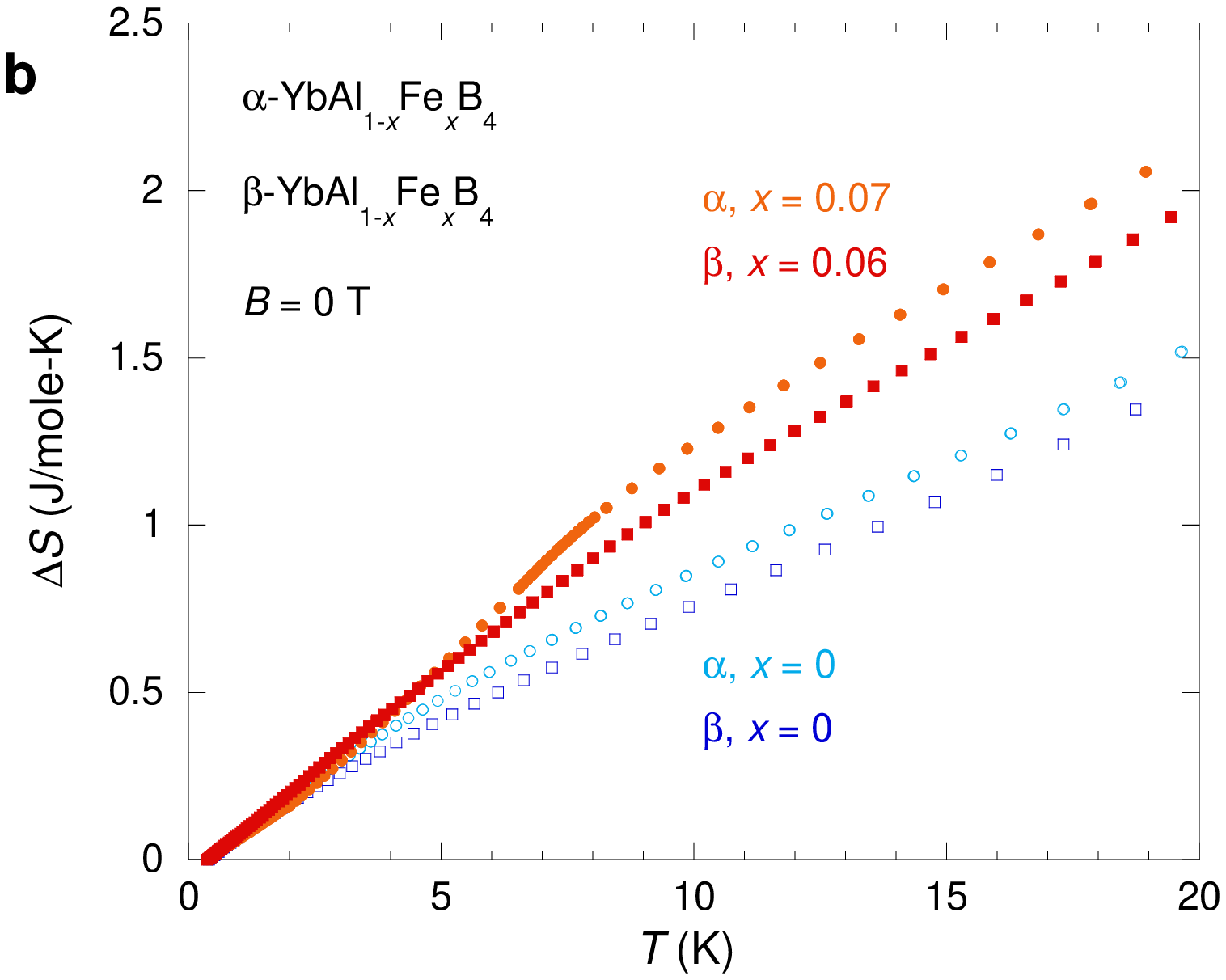}
\end{minipage}\hspace{2pc}%
\caption{Temperature dependence of (a) the specific heat divided by temperature $C_P/T$ and (b) the entropy of $\alpha$-YbAlB$_4$ (open circle), $\alpha$-YbAl$_{0.93}$Fe$_{0.07}$B$_4$ (closed circle), $\beta$-YbAlB$_4$ (open square), $\beta$-YbAl$_{0.94}$Fe$_{0.06}$B$_4$ (closed square), and $\beta$-YbAl$_{0.95}$Fe$_{0.05}$B$_4$ (closed triangle). The panel (a) inset is the temperature dependence of $C_P/T$ near the transition temperature under 0 T (closed circle for $\alpha$-YbAl$_{0.93}$Fe$_{0.07}$B$_4$ and closed square for $\beta$-YbAl$_{0.94}$Fe$_{0.06}$B$_4$) and 9 T (open circle for $\alpha$-YbAl$_{0.93}$Fe$_{0.07}$B$_4$ and open square for $\beta$-YbAl$_{0.94}$Fe$_{0.06}$B$_4$) along $c$-axis.}
\end{figure*}

Figures 1(a) and (b) show the temperature dependence of the susceptibility $\chi = M/H$ for both $\alpha$-YbAl$_{1-x}$Fe$_x$B$_4$ and $\beta$-YbAl$_{1-x}$Fe$_x$B$_4$, respectively. The susceptibility for both systems is clearly Ising like. Namely, the $c$-axis susceptibility is strongly temperature dependent, while the $ab$-plane component is nearly temperature independent with a small value of $\sim 0.005$ emu/mol. The $c$-axis component for $\alpha$-YbAlB$_4$ exhibits no anomaly down to 2 K, and start leveling off below $T^* = 8$ K, indicating the onset of the Fermi liquid ground state \cite{Matsumoto1,Matsumoto2}. In contrast, the $c$-axis susceptibility  for $\beta$-YbAlB$_4$ exhibits divergent behavior on cooling, reflecting the unconventional quantum criticality \cite{Nakatsuji,Matsumoto1}. With doping of Fe by 3\%, the $c$-axis susceptibility for $\beta$-YbAl$_{0.97}$Fe$_{0.03}$B$_4$ exhibits a weaker temperature dependence with a smaller value than the pure case below $\sim 20$ K, while it completely overlaps that for the pure $\beta$-YbAlB$_4$ at $T > 20$ K. In contrast, no change was found in the $ab$-plane component by doping of Fe by 3\% for the $\beta$ phase (Fig. 1(b)).

With further substitution of Fe at the Al site, however, both $\alpha$-YbAl$_{0.93}$Fe$_{0.07}$B$_4$ and $\beta$-YbAl$_{0.94}$Fe$_{0.06}$B$_4$ exhibit a weak kink in the temperature dependence of the $c$-axis susceptibility at 7.5(5) K and 9.5(5) K, respectively and bifurcate into different curves for zero-field cooled and field cooled sequences. More clear anomaly and hysteresis was found in the temperature dependence of the $ab$-plane susceptibility, suggesting the ferromagnetic component lies in the $ab$-plane. The insets of Figures 1(a) and (b) show the field dependence of the magnetization of $\alpha$-YbAl$_{0.93}$Fe$_{0.07}$B$_4$ and $\beta$-YbAl$_{0.94}$Fe$_{0.06}$B$_4$ under a field along $ab$-plane at 2 K. For each measurement, the sample was first cooled down to 2 K from the paramagnetic state under zero field. Then, the field was increased up to 7 T and decreased back to zero to obtain a magnetization curve. Each sample shows an almost linear magnetic field dependence of the magnetization and no hysteresis was found within experimental resolution of $1 \times 10^{-4} \mu_{\rm B}/$Yb, which places the upper bound for the ferromagnetic component along $ab$-plane. Both linear magnetization curve and the small size of the spontaneous moment point to a canted antiferromagnetism. On the other hand, the temperature dependence of the susceptibility for the Lu analog $\alpha$-LuAl$_{0.79}$Fe$_{0.21}$B$_4$ and $\alpha$-LuAl$_{0.96}$Fe$_{0.04}$B$_4$ shown in Figs. 1(a) and (b)~shows diamagnetism with a small negative value in between $-1 \times 10^{-4}$ to $-3 \times 10^{-5}$ emu/mol, indicating that Fe ion is non-magnetic and thus it is 4$f$ moments of Yb ion that form the magnetic order in both $\alpha$-YbAl$_{0.93}$Fe$_{0.07}$B$_4$ and $\beta$-YbAl$_{0.94}$Fe$_{0.06}$B$_4$.

Figure 2 presents the temperature dependence of the specific heat divided by temperature $C_P/T$ at 0 T. Both pure phases show paramagnetic behavior down to the lowest temperatures. Namely, $C_P/T$ for $\alpha$-YbAlB$_4$ gradually increases on cooling and saturates to a large value of around 130 mJ/mol K$^2$, indicating the formation of a heavy Fermi liquid state \cite{Matsumoto1,Matsumoto2}. $C_P/T$ for $\beta$-YbAlB$_4$ shows a logarithmic divergence, consistent with the non-Fermi liquid ground state \cite{Nakatsuji,Matsumoto1}. 

On the other hand, the doping of Fe induces the anomalies in $C_P/T$ due to the magnetic transitions: $\alpha$-YbAl$_{0.93}$Fe$_{0.07}$B$_4$ exhibits a peak at 6.7(3) K and $\beta$-YbAl$_{0.95}$Fe$_{0.05}$B$_4$ and $\beta$-YbAl$_{0.94}$Fe$_{0.06}$B$_4$ show a shoulder like anomaly at 8.3 K. Given the tails of the anomalies with a temperature width of $\sim 1$ K, these confirm the bulk nature of the magnetic transition inferred from the above susceptibility measurements. The inset of Fig. 2 presents the temperature dependence of $C_P/T$ under fields of 0 T and 9 T along the $c$-axis. The application of a magnetic field of 9 T slightly decreases the peak temperature of $C_P/T$ by $\sim 0.5$ K, indicating that the magnetic ordered state is not ferromagnetic but antiferromagnetic. On further cooling, all the doped samples show an additional anomaly at around 3 K, suggesting another magnetic transition, which is not seen in the temperature dependence of the susceptibility. For $\beta$-YbAl$_{1-x}$Fe$_{x}$B$_{4}$, the fact that both low and high temperature anomalies appear at nearly the same temperatures for both $x = 0.05$ and 0.06 indicates that both transitions are intrinsic and not due to the mosaicity found for $x$ = 0.06. One possible origin of the low temperature anomaly is the change of the magnetic structure. Further microscopic studies such as neutron diffraction and nuclear magnetic resonance measurements are necessary to determine the spin structure.

In order to gain the insight of the origin of the magnetic order induced by Fe substitution, we made the Curie-Weiss (CW) analysis for the susceptibility. For both pure $\alpha$- and $\beta$-phases as well as $\beta$-YbAl$_{0.97}$Fe$_{0.03}$B$_4$, the susceptibility above 20 K collapses on top of each other and the CW fitting at $T >$ 150 K yields the effective moment $P_{\rm eff}$ of 2.2(2) $\mu_{\rm B}$ and the antiferromagnetic (AFM) Weiss temperature $\Theta_{\rm W}$ of 110(5) K \cite{Matsumoto1,Matsumoto2}. On the other hand, the same fitting at $T >$ 150 K for $\alpha$-YbAl$_{0.93}$Fe$_{0.07}$B$_4$ and $\beta$-YbAl$_{0.94}$Fe$_{0.06}$B$_4$ respectively yields $P_{\rm eff}$ = 2.44 $\mu_{\rm B}$ and AFM $\Theta_{\rm W}$ = 60 K, and $P_{\rm eff}$ = 2.2(2) $\mu_{\rm B}$ and AFM $\Theta_{\rm W}$ = 80 K. The analyses indicate that the Fe substitution reduces the Weiss temperature by $30 \sim 40$\%. This suggests that the Kondo temperature, which is estimated to be $\sim 200$ K for both pure $\alpha$ and $\beta$ phases  \cite{Nakatsuji,Matsumoto1,Matsumoto2}, becomes significantly suppressed by Fe doping. This is consistent with the chemical pressure effect inferred from the crystal structure analysis, as the pressure normally renders the Yb system more magnetic.

Correspondingly, the entropy for both phases estimated by integrating $C_P/T$ from the lowest $T$ = 0.4 K (Fig. 2(b)) indicates a substantial increase at 20 K with the Fe substitution of $6 \sim 7$ \%. In both pure systems, the ground state of the crystal electric field scheme is known to have a Kramers doublet, which is most likely separated by more than 100 K from the exited doublet state \cite{Andriy,Matsumoto2}. Assuming that the gap scale stays the same by Fe doping, we may conclude that the increase of the entropy at 20 K indicates again the suppression of the Kondo temperature. Moreover, even at the magnetic transition temperatures $\sim 9$ K, the entropy of both $\alpha$-YbAl$_{0.93}$Fe$_{0.07}$B$_4$ and $\beta$-YbAl$_{0.94}$Fe$_{0.06}$B$_4$  is larger than the pure systems by $\sim 0.3$ mJ/mol-K. This value is too large for a pure ferromagnetic transition given the spontaneous moment $< 0.001 \mu_{\rm B}$/Yb estimated by the magnetization measurement. This also provides another evidence that the observed hysteresis is due to a canted antiferromagnetism, not by a simple ferromagnetism.

To summarize, we found that the Fe substitution at Al site induces a magnetic order, most likely a canted antiferromagnetic type due to 4$f$ electrons, by suppressing the Kondo temperature. The crystal structure analysis indicates that the Fe substitution decreases the volume without having any structural transition. Combined, the magnetic order is induced in both $\alpha$-YbAlB$_4$ and $\beta$-YbAlB$_4$ because of the chemical pressure applied by the Fe substitution and thus indicates the proximity to a magnetic instability for both phases of YbAlB$_4$ at ambient pressure. The detailed study for the ground state evolution from pure $\alpha$- and $\beta$-YbAlB$_4$ with more fine steps of Fe doping is necessary to clarify how a magnetic quantum criticality emerges and develops with doping Fe in both $\alpha$- and $\beta$-phases. It is also an interesting future issue how a putative magnetic quantum criticality induced by the Fe substitution is related with the unconventional quantum criticality observed in the heavy fermion superconductor $\beta$-YbAlB$_4$.

We thank K. Sone, Y. Matsumoto, E.C.T. O'Farrell, and T. Tomita for useful discussions.
This work was partially supported by Grants-in-Aid (No.21684019) from 
JSPS, by Grants-in-Aids for Scientific Research on Innovative Areas ``Heavy Electrons" of MEXT, Japan and Toray Science Foundation. 
J.Y.C. would like to acknowledge National Science Foundation (NSF) DMR1063735 for partial support of this project.



\begin{thebibliography}{99}
\bibitem{Vojta} H. v. L\"{o}hneysen, A. Rosch, M. Vojta, and P. W\"{o}lfle, Rev. Mod. Phys. {\bf 79}, 1015 (2007).
\bibitem{Monthoux} P. Monthoux, D. Pines, and G. G. Lonzarich, Nature {\bf 450}, 1177 (2007).
\bibitem{Gegenwart} P. Gegenwart, Q. Si, and F. Steglich, Nature Phys. {\bf 4}, 186 (2008).
\bibitem{Lohneysen} H. v. Lohneysen, T. Pietrus, G. Portisch, H. G. Schlager, A. Schroder, M. Sieck, T. Trappmann, Phys. Rev. Lett. {\bf 72}, 3262 (1994).
\bibitem{Trovarelli} O. Trovarelli, C. Geibel, S. Mederle, C. Langhammer, F. M. Grosche, P. Gegenwart, M. Lang, G. Sparn, and F. Steglich, Phys. Rev. Lett. {\bf 85}, 626 (2000).
\bibitem{Nakatsuji} S. Nakatsuji, K. Kuga, Y. Machida, T. Tayama, T. Sakakibara, Y. Karaki, H. Ishimoto, S. Yonezawa, Y. Maeno, E. Pearson, G. G. Lonzarich, L. Balicas, H. Lee, and Z. Fisk, Nature Phys. {\bf 4} 603 (2008).
\bibitem{Moriya} T. Moriya, {\it Spin Fluctuations in Itinerant Electron Magnetism} (Springer, Berlin, 1985).
\bibitem{Millis} A. J. Millis, Phys. Rev. B {\bf48}, 7183 (1993).
\bibitem{Macaluso} R. T. Macaluso, S. Nakatsuji, K. Kuga, E. L. Thomas, Y. Machida, Y. Maeno, Z. Fisk, J. Y. Chan, Chem. Mater. {\bf19}, 1918 (2007). 
\bibitem{Kuga} K. Kuga, Y. Karaki, Y. Matsumoto, Y. Machida, and S. Nakatsuji, Phys. Rev. Lett. {\bf 101} 137004 (2008).
\bibitem{Matsumoto1} Y. Matsumoto, S. Nakatsuji, K. Kuga, Y. Karaki, N. Horie, Y. Shimura, T. Sakakibara, A. H. Nevidomskyy, and P. Coleman, Science {\bf 21} 316 (2011).
\bibitem{Okawa} M. Okawa, M. Matsunami, K. Ishizaka, R. Eguchi, M. Taguchi, A. Chainani, Y. Takata, M. Yabashi, K. Tamasaku, Y. Nishino, T. Ishikawa, K. Kuga, N. Horie, S. Nakatsuji, and S. Shin, Phys. Rev. Lett. {\bf 104} 247201 (2010).
\bibitem{eoinPRL}E. C. T. O'Farrell, D. A. Tompsett, S. E. Sebastian, N. Harrison, C. Capan, L. Balicas, K. Kuga, A. Matsuo, K. Kindo, M. Tokunaga, S. Nakatsuji, G. Cs\'anyi, Z. Fisk, and M. L. Sutherland, Phys. Rev. Lett. {\bf 102}, 216402 (2009).
\bibitem{Matsumoto2} Y. Matsumoto, K. Kuga, T. Tomita, Y. Karaki, and S. Nakatsuji, Phys. Rev. B {\bf 84}, 125126 (2011).
\bibitem{Holanda} L. M. Holanda, J. M. Vargas, W. Iwamoto, C. Rettori, S. Nakatsuji, K. Kuga, Z. Fisk, S. B. Oseroff, and P. G. Pagliuso, Phys. Rev. Lett. {\bf 107}, 026402 (2011).
\bibitem{eoinPRL2}E. C. T. O'Farrell, Y. Matsumoto, S. Nakatsuji, Phys. Rev. Lett. {\bf 109}, 176405 (2012).
\bibitem{Altomare} A. Altomare, M. C. Burla, M. Camalli, G. L. Cascarano, C. Giacovazzo, A. Guagliardi, A. G. G. Moliterni, G. Polidori, and R. Spagna, J. Appl. Crystallogr. {\bf 32}, 115 (1999).
\bibitem{Sheldrick} G. M. Sheldrick, Acta Cryst. {\bf A64}, 112 (2008).
\bibitem{Andriy} A. H. Nevidomskyy and P. Coleman, Phys. Rev. Lett. {\bf 102} 077202 (2009).
\end{thebibliography}
\end{document}